\documentclass[eqsecnum,nofootinbib,12pt]{revtex4}
\usepackage{amsmath}
\usepackage{amsthm,amscd}
\usepackage{mathrsfs}
\usepackage{verbatim}
\usepackage{amsfonts,amsmath,latexsym,amssymb,txfonts,bm}
\usepackage[american]{babel}

\begin{document}
\title{Non-Perturbative One-Loop Effective Action for Electrodynamics in Curved Spacetime}
\author{
Guglielmo Fucci \footnote{Electronic address: gfucci@nmt.edu and Guglielmo\textunderscore Fucci@baylor.edu}
\thanks{Electronic address: gfucci@nmt.edu}}
\altaffiliation[Also at ]{Department of Mathematics, Baylor University One Bear Place 9653, Waco, TX USA}
\affiliation{Department of Physics, New Mexico Institute of Mining and Technology,
Socorro, NM 87807 USA
}
\date{\today}
\vspace{2cm}
\begin{abstract}
In this paper we explicitly evaluate the one-loop effective action in four
dimensions for scalar and spinor fields under the influence of a strong,
covariantly constant, magnetic field in curved spacetime. In the framework
of zeta function regularization, we find the one-loop effective action to
all orders in the magnetic field up to linear terms in the Riemannian curvature.
As a particular case, we also obtain the one-loop effective action for massless
scalar and spinor fields. In this setting, we found that the vacuum energy of
charged spinors with small mass becomes very large due entirely by the gravitational correction.
\end{abstract}
\maketitle
\section{Introduction}
It is well known that the effective action plays a major and important role in
quantum field theory and quantum gravity \cite{schwinger51,dewitt65,dewitt03}.
In particular, the knowledge of the effective action enables one to obtain the
relevant full one-point propagators and the full vertex functions which, in turn,
constitute the building blocks of the $S$-matrix \cite{dewitt65}. Moreover,
the effective action, upon variation of the independent fields, gives the effective
dynamical equations which describe the back-reaction of the quantum fields on the
classical background.
One of the most effective mathematical tools to study the propagators and the
effective action in quantum field theory is the proper time method or heat kernel method which
was developed by Schwinger in \cite{schwinger51,schwinger54} and then generalized
to include curved spacetime by DeWitt in \cite{dewitt67a,dewitt75}. Some nice
reviews on this subject can be found in \cite{barvinsky85,vassile03}.

One of the most important achievements of quantum field theory in the past century was
the development of quantum electrodynamics. In particular, Schwinger \cite{schwinger51}
used the heat kernel method to obtain the effective action for constant electromagnetic fields.
Moreover, he showed that, in presence of an electric field, the effective action
acquires an imaginary part which was interpreted as probability of production of
electron-positron pairs induced by the electric field. This is known as Schwinger mechanism
which becomes exponentially small for weak electric fields. In recent years a generalization
of Schwinger results for the effective action and pair production have been obtained
for inhomogeneous background electromagnetic fields \cite{dunne99,dunne04,dunne05}. Moreover,
the effective action for Dirac spinors under the influence of general uniform magnetic fields
was computed, in the massive and massless case, in \cite{soldati98}.

Although of extreme interest, the above results were obtained in flat (Minkowski) spacetime.
A great amount of work has been done in order to study the case in which both the
electromagnetic and gravitational fields are present. Of course, this is not a trivial task and
different approximation schemes have been used in order to obtain relevant information in
different regimes \cite{avramidi94,avramidi00,avramidi02,vassile03,barvinsky85}.
In this paper, we are interested in the situation in which the
electromagnetic field is much stronger than the gravitational field. In this case, the
electromagnetic field cannot be treated as a perturbation and, therefore, essentially
non-perturbative techniques need to be exploited. A powerful and promising non-perturbative algebraic
approach, combining gauge fields and gravitational fields, was developed in \cite{avramidi08,avramidi08a}
where homogeneous bundles with parallel curvature over symmetric spaces were considered.

Recently, in \cite{avramidi08e}, the heat kernel (which in turn leads to the effective action)
for a Laplace type operator on parallel homogeneous Abelian bundles for arbitrary
Riemmanian manifolds was computed. The first three coefficients of the heat kernel asymptotic
expansion in powers of the Riemannian curvature but to all orders of the electromagnetic field were evaluated.
The results obtained in \cite{avramidi08e} were applied to the computation of the contribution
to the imaginary part of the effective action for scalar and spinor fields in a strong electromagnetic field
obtaining a generalization of the original Schwinger result to curved spacetime \cite{avramidi09a}.
A non-perturbative effective action in the electromagnetic field in arbitrary
curved spacetime for scalar and spinor fields was recently obtained in the framework
of worldline formalism in \cite{bastianelli09}.

In this paper we will utilize the non-perturbative results obtained in \cite{avramidi08e,avramidi09a}
in order to compute the one-loop effective action for scalar and spinor fields under the influence of
a strong magnetic field on an arbitrary curved spacetime up to linear terms in the Riemannian curvature.
We will assume, throughout the paper, that the electric field vanishes identically. Under this assumption
the effective action is purely real (the heat kernel becomes an entire function) and effects of creation of pairs
do not occur. 
By using the non-perturbative heat kernel asymptotic expansion developed in \cite{avramidi09a},
we will find, through the Mellin transform, the spectral $\zeta$-function and, hence, the one-loop effective action.

We would like to stress, at this point, that the non-perturbative one-loop effective action
is found, here, by using $\zeta$-function regularization method which is a different approach from the
worldline formalism considered in \cite{bastianelli09}. Obviously, the two methods should lead to
comparable results.

In the rest of this paper we will use the standard convention of setting $\hbar=c=1$.

\section{Effective action and  $\zeta$-function }

Let us consider a smooth compact $n$-dimensional Riemannian manifold $\cal{M}$ without boundary
endowed with a positive-definite metric $g_{\mu\nu}$. Let $\varphi$ be smooth sections
of a complex spin bundle $\cal{S}$, with structure group
${\rm Spin}(n)\times U(1)$, over the manifold $\cal{M}$ . Let us denote by $\nabla$ the total connection
of the bundle $\cal{S}$ containing the spin connection as well as the Riamannian connection.
Obviously, the commutator of the covariant derivatives defines the curvature on $\cal{S}$
\begin{equation}\label{0}
[\nabla_\mu,\nabla_\nu]\varphi=({\cal R}_{\mu\nu}+i F_{\mu\nu})\varphi\;,
\end{equation}
where $F_{\mu\nu}$ represents the curvature of the $U(1)$ connection, describing the
magnetic field, and $\cal{R}_{\mu\nu}$ is the curvature of the spin connection
which for scalar fields vanishes identically
\begin{equation}
{\cal R}_{\mu\nu}=0\;,
\end{equation}
and for spinor fields it has the form
\begin{equation}
{\cal R}_{\mu\nu}=\frac{1}{4}R^{ab}{}_{\mu\nu}\gamma_{[a}\gamma_{b]}\,.
\end{equation}
Here, $R^{ab}{}_{\mu\nu}$ represents the Riemann tensor and $\gamma_{a}$ are the Dirac matrices which satisfy the
Clifford algebra
\begin{equation}
\gamma_{a}\gamma_{b}+\gamma_{b}\gamma_{a}=2g_{ab}\mathbb{I}\;.
\end{equation}

Let $\mathscr{L}$ be a Laplace type second order partial differential operator acting on smooth sections
of the bundle $\cal{S}$. In general we write the operator $\mathscr{L}$ as
\begin{equation}\label{1}
\mathscr{L}=-\Delta+\xi R+Q\;,
\end{equation}
where $\Delta=g^{\mu\nu}\nabla_{\mu}\nabla_{\nu}$, $\xi$ is a constant parameter, $R$ is the scalar curvature
and $Q$ is a smooth endomorphism of the bundle $\cal{S}$. This operator is elliptic self-adjoint and it has a positive definite leading symbol \cite{gilkey95}.

For the scalar fields case, there are usually two interesting choices for the parameter $\xi$.
In the minimally coupled case one sets $\xi^{\rm scalar}=0$ and in the conformally coupled case one sets $\xi^{\rm scalar}=(n-2)/4(n-1)$.
Since we do not consider self-interacting scalar fields, we set the endomorphism $Q^{\rm scalar}=0$.

For the spinor fields case, (\ref{1}) is obtained by taking the square of the Dirac operator $D=i\gamma^{\mu}\nabla_{\mu}$.
It is not difficult to show that the operator (\ref{1}) takes the form
\begin{equation}
\mathscr{L}^{\rm spinor}=-\Delta+\frac{1}{4}R-\frac{1}{2}iF_{\mu\nu}\gamma^{[\mu}\gamma^{\nu]}\,.
\end{equation}

It is well known that the Euclidean one-loop effective action is given as functional
determinant of the operator of small disturbance as
\begin{equation}\label{2}
\Gamma_{(1)}=\sigma\log\,{\rm Det}\left(\frac{\mathscr{L}+M^{2}}{\mu^{2}}\right)\;,
\end{equation}
where $M$ is the mass of the field which is assumed to be large enough to ensure the positivity of the spectrum
of the operator $\mathscr{L}$, $\mu$ is an additional mass parameter which should be fixed by renormalization and is necessary in order to make the argument of the logarithm dimensionless, and finally $\sigma$ is the fermion number which is $(-1)$ for spinor fields
and $(+1)$ for charged scalar fields \cite{hawking77} (we consider, here, complex fields). The functional determinant (\ref{2}), however, needs to be regularized.
We will use in this paper the $\zeta$-function regularization, of course this is not the only way to
define the above functional determinant (see e.g. \cite{cognola93}).

The spectral $\zeta$-function, $\zeta(s)$,  for a positive operator $\mathscr{L}$ can be analytically continued
to a meromorphic function in the entire complex plane with only simple poles (residues of which are given by the heat kernel coefficients of $\mathscr{L}$) and regular at $s=0$ \cite{kirsten01}.
An important relation exists between the spectral $\zeta$-function of $\mathscr{L}$ and its trace of the heat kernel diagonal.
By recalling the integral representation of the gamma function, it is not difficult
to show that the $\zeta$-function is obtained by inverse Mellin transform as follows \cite{dewitt03,elizalde94,kirsten01,vassile03}
\begin{equation}\label{3}
\zeta(s)=\frac{\mu^{2s}}{\Gamma(s)}\int_{0}^{\infty}dt\,t^{s-1}{\rm Tr}\,\exp(-t\mathscr{L})\;,
\end{equation}
where $\Gamma(s)$ denotes the Euler gamma function and, for convenience, we have included the mass parameter $\mu$.
The one-loop effective action (\ref{2}) is
expressed in terms of the $\zeta$-function, and hence in terms of the trace of the heat kernel diagonal via (\ref{3}),
as \cite{dewitt03,elizalde94,kirsten01,vassile03}
\begin{equation}
\Gamma_{(1)}=-\sigma\zeta^{\prime}(0)\;,
\end{equation}
where $\zeta^{\prime}(0)$ denotes the derivative of $\zeta(s)$ evaluated at $s=0$.

In this paper, we will use the representation of the $\zeta$-function (\ref{3}) and
the non-perturbative heat kernel asymptotic expansion developed in \cite{avramidi08e,avramidi09a}
to evaluate the one-loop effective action for scalar and spinor fields
under the influence of a strong magnetic field in curved spacetime.

\section{Heat Kernel Asymptotic Expansion}

In order to present the main results for the asymptotic expansion of the
heat kernel diagonal, we will briefly describe the spectral decomposition
of the $U(1)$ curvature $F_{\mu\nu}$ following \cite{avramidi09a}.

For the matrix $F=(F^{\mu}{}_{\nu})$ we have the following decomposition
\begin{equation}
F=\sum_{k=1}^{N}B_{k}E_{k}\;,
\end{equation}
where $B_{k}$ are real invariants, that we will call magnetic fields, $N\leq[n/2]$, and $E_{k}$ are antisymmetric matrices which satisfy the relations
\begin{equation}
E^{k}_{\mu[\nu}E^{k}_{\alpha\beta]}=0\;,\quad\textrm{and}\quad E_{k}E_{m}=0\; \quad\textrm{for}\;k\neq m \;.
\end{equation}
Moreover, in the Euclidean setting, we introduce the projectors onto $2$-dimensional eigenspaces, $\Pi^{k}$, which
are symmetric matrices defined as
\begin{equation}
\Pi_{k}=-E_{k}^{2}\;.
\end{equation}
Obviously, they satisfy the relations
\begin{equation}
\Pi_{k}^{2}=\Pi_{k}\;,\quad E_{k}\Pi_{k}=\Pi_{k}E_{k}=E_{k}\;,\quad \Pi_{k}^{\mu}{}_{\mu}=2\;,
\end{equation}
and for $k\neq m$,
\begin{equation}
E_k \Pi_m=\Pi_m E_k=0\,, \qquad
\Pi_k\Pi_m=0\,.
\end{equation}
In what follows, we will assume that all the invariants are the same, namely $B_1=\cdots=B_N=B$.

In the paper \cite{avramidi09a}, it was shown that the trace of the heat kernel diagonal, up to
linear terms in the Riemannian curvature, for a covariantly constant magnetic field, has the following asymptotic expansion for scalar fields
\begin{equation}\label{4}
{\rm Tr}\,\exp\{-t\mathscr{L}^{\rm scalar}\}\sim
(4\pi t)^{-n/2}\Phi^{\rm scalar}(t)\left\{1+t\mathcal{B}^{\rm scalar}_1(t)
+\cdots\right\}\,,
\end{equation}
where
\begin{equation}\label{4a}
\Phi^{\rm scalar}(t)=
\left(\frac{tB}{\sinh(tB)}\right)^N\,,
\end{equation}
\begin{eqnarray}\label{4b}
\mathcal{B}^{\rm scalar}_1(t)&=&
\left(\frac{1}{6}-\xi\right)R
+\varphi(t)H^{\mu\nu}_1 R_{\mu\nu}
+\rho(t)H^{\mu\nu}_1H^{\alpha\beta}_1R_{\mu\alpha\nu\beta}
-\sigma(t)X^{\mu\alpha}_1X^{\nu\beta}_1 R_{\mu\alpha\nu\beta}\,,
\end{eqnarray}
with
\begin{equation}
H^{\mu\nu}_1=\sum_{k=1}^{N}\Pi_{k}^{\mu\nu}\,,
\qquad
X^{\mu\nu}_1=\sum_{k=1}^{N}E_{k}^{\mu\nu}\,,
\end{equation}
\begin{eqnarray}\label{4c}
\varphi(t)&=&
\frac{1}{6}
+\frac{3}{8}\frac{1}{(tB)^2}
-\frac{1}{24}tB\coth(tB)
-\frac{3}{8}\frac{\coth(tB)}{tB}\,,
\\[5pt]
\sigma(t)&=&
\frac{1}{16}
-\frac{3}{32}\frac{\coth(tB)}{tB}
+\frac{3}{32}\frac{1}{\sinh^2(tB)}
\\[5pt]
\rho(t)\label{4d}
&=&
-\frac{5}{24}
-\frac{3}{8}\frac{1}{(tB)^2}
+\frac{1}{24}tB\coth(tB)
+\frac{7}{16}\frac{\coth(tB)}{tB}
-\frac{1}{16}\frac{1}{\sinh^2(tB)}\,.
\end{eqnarray}
Here, and for the rest of the paper, $R$ denotes the scalar curvature while $R_{\mu\nu}$ and $R_{\mu\nu\alpha\beta}$ denote, respectively,
the Ricci and Riemann tensors.

For spinor fields, one obtains an asymptotic expansion for the trace of the heat kernel diagonal
similar to the one in (\ref{4}). In this case, one can write, up to linear terms in the Riemannian curvature, \cite{avramidi09a}
\begin{equation}\label{4e}
{\rm Tr}\,\exp\{-t\mathscr{L}^{\rm spinor}\}\sim
(4\pi t)^{-n/2}\Phi^{\rm spinor}(t)\left\{1+t\mathcal{B}^{\rm spinor}_1(t)
+\cdots\right\}\,,
\end{equation}
where
\begin{equation}\label{4f}
\Phi^{\rm spinor}(t)=2^{[n/2]}\left[tB\coth(tB)\right]^N\,,
\end{equation}
\begin{eqnarray}\label{4f1}
\mathcal{B}^{\rm spinor}_1(t)&=&
-\frac{1}{12}R
+\varphi(t)H^{\mu\nu}_1R_{\mu\nu}
+\rho(t)H^{\mu\nu}_1 H^{\alpha\beta}_1R_{\mu\alpha\nu\beta}
-\lambda(t)X^{\mu\alpha}_1 X^{\nu\beta}_1
R_{\mu\alpha\nu\beta}
\,,
\end{eqnarray}
and
\begin{eqnarray}\label{4g}
\lambda(t)&=&
-\frac{3}{16}
+\frac{3}{32}\frac{1}{\sinh^2(tB)}
+\frac{1}{4}\frac{\tanh(tB)}{tB}
-\frac{3}{32}\frac{\coth(tB)}{tB}
\,.
\end{eqnarray}

It is clear, at this point, that in both scalar and spinor cases the spectral $\zeta$-function
for the operator $(\mathscr{L}+M^{2})$ is given, in the approximation under consideration, by
\begin{equation}
\zeta(s)=\int_{\mathcal{M}}d{\rm vol}\,\zeta_{\rm loc}(s)\;,
\end{equation}
where the local spectral $\zeta$-function is
\begin{equation}\label{5}
\zeta_{\rm loc}(s)=\frac{\mu^{2s}}{(4\pi)^{n/2}\Gamma(s)}\int_{0}^{\infty}dt\,t^{s-n/2-1}e^{-tM^{2}}\Phi(t)\left\{1+t\mathcal{B}_1(t)\right\}\;.
\end{equation}
Since the one-loop effective action can be written in terms of the effective Lagrangian, $\mathcal{L}$, as
\begin{equation}\label{5ab}
\Gamma_{(1)}=\int_{\mathcal{M}}d{\rm vol}\,\mathcal{L}\;,
\end{equation}
one soon realizes that
\begin{equation}
\mathcal{L}=-\sigma\zeta_{\rm loc}^{\prime}(0)\;.
\end{equation}
Because of the relation (\ref{5ab}), in the rest of this paper we will be mainly focused on the evaluation of the effective Lagrangian
for both scalar and spinor fields.

The next task is, thus, the computation of the integral (\ref{5}) to obtain an explicit expression for
$\zeta_{\rm loc}(s)$.

\section{Integrals}

In this section, we will derive a general formula which will enable us to
evaluate the function $\zeta_{\rm loc}(s)$ in (\ref{5}). Let $m$ and $q$ be non-negative integers, then the integrals that we will need to compute
have the following general form
\begin{equation}\label{5a}
I_{m,q}(\alpha,M^{2},B)=\int_{0}^{\infty}dt\,t^{\alpha-1}e^{-tM^{2}}\frac{\cosh^{m}(tB)}{\sinh^{q}(tB)}\;,
\end{equation}
which are well defined for ${\rm Re}M^{2}>0$ and ${\rm Re}\,\alpha>(q+1)$. It is convenient to
rewrite the power of $\cosh(tB)$ in terms of functions of multiples of the argument \cite{gradshtein07} to get
\begin{eqnarray}\label{6}
I_{2m,q}(\alpha,M^{2},B)&=&\frac{1}{2^{2m}}\sum_{k=0}^{m-1}{2m \choose k}\int_{0}^{\infty}dt\,\frac{t^{\alpha-1}}{\sinh^{q}(tB)}\Big[e^{-t[M^{2}-2(m-k)B]}+e^{-t[M^{2}+2(m-k)B]}\Big]\nonumber\\
&+&\frac{1}{2^{2m}}{2m \choose m}\int_{0}^{\infty}dt\,t^{\alpha-1}\frac{e^{-tM^{2}}}{\sinh^{q}(tB)}\;,
\end{eqnarray}
and
\begin{eqnarray}\label{7}
I_{2m-1,q}(\alpha,M^{2},B)=\frac{1}{2^{2m-1}}\sum_{k=0}^{m-1}{2m-1 \choose k}\int_{0}^{\infty}dt\,\frac{t^{\alpha-1}}{\sinh^{q}(tB)}\Big[e^{-t[M^{2}-(2m-2k-1)B]}+e^{-t[M^{2}+(2m-2k-1)B]}\Big]\;.\nonumber\\
\end{eqnarray}

By inspection of the expressions above, it is not difficult to realize that
the integrals (\ref{6}) and (\ref{7}) have been reduced to a linear combination of integrals of the
following type
\begin{equation}\label{7a}
\mathcal{I}_{q}(\alpha,M^{2},B)=\int_{0}^{\infty}dt\,t^{\alpha-1}\frac{e^{-tM^{2}}}{\sinh^{q}(tB)}\;.
\end{equation}
The integral $\mathcal{I}_{q}$ can be readily evaluated by by using the formula \cite{gradshtein07,avramidi99,erdelyi53}
\begin{equation}\label{8}
\int_{0}^{\infty}dt\,t^{\alpha-1}\frac{e^{-\gamma t}}{1-\lambda e^{-t}}=\Gamma(\alpha)\Phi(\lambda,\alpha,\gamma)\;,
\end{equation}
which is well defined for ${\rm Re}\gamma>0$ and either $\lambda\leq 1$,
$\lambda\neq 1$, ${\rm Re}s>0$ or $\lambda=1$ and ${\rm Re}s>1$.
In (\ref{8}), $\Phi$ is the Lerch transcendent, which is a generalization
of the Hurwitz $\zeta$-function, defined as \cite{gradshtein07,erdelyi53}
\begin{equation}\label{8a}
\Phi(\lambda,\alpha,\gamma)=\sum_{n=0}^{\infty}\frac{\lambda^{n}}{(n+\gamma)^{\alpha}}\;.
\end{equation}

By repeatedly differentiating the equation (\ref{8}) with respect to $\lambda$, we obtain an
expression for the integral (\ref{7a}), more explicitly
\begin{equation}\label{9}
\mathcal{I}_{q}(\alpha,M^{2},B)=B^{-\alpha}\frac{2^{q-\alpha}\Gamma(\alpha)}{\Gamma(q)}
\frac{\partial^{q-1}}{\partial\lambda^{q-1}}\left[\Phi\left(\lambda,\alpha,\frac{M^{2}-qB}{2B}+1\right)\right]\Bigg|_{\lambda=1}\;.
\end{equation}
By exploiting the result (\ref{9}) and by introducing the dimensionless variable $z=M^{2}/2B$, we write the integrals (\ref{5a}) as follows
\begin{equation}\label{10}
I_{m,q}(\alpha,M^{2},B)=B^{-\alpha}f_{m,q}(\alpha,z)\;,
\end{equation}
where the functions $f_{m,q}(\alpha,z)$ have the expressions
\begin{eqnarray}\label{10a}
f_{2m,q}(\alpha,z)&=&\frac{2^{q-\alpha-2m}\Gamma(\alpha)}{\Gamma(q)}\sum_{k=0}^{m-1}{2m \choose k}\frac{\partial^{q-1}}{\partial\lambda^{q-1}}\Bigg[\Phi\left(\lambda,\alpha,z-m+k-\frac{q}{2}+1\right)\nonumber\\
&+&\Phi\left(\lambda,\alpha,z+m-k-\frac{q}{2}+1\right)+{2m\choose m}\,\Phi\left(\lambda,\alpha,z-\frac{q}{2}+1\right)\Bigg]\Bigg|_{\lambda=1}\;,
\end{eqnarray}
\begin{eqnarray}\label{10b}
f_{2m-1,q}(\alpha,z)&=&\frac{2^{q-\alpha-2m+1}\Gamma(\alpha)}{\Gamma(q)}\sum_{k=0}^{m-1}{2m-1 \choose k}\frac{\partial^{q-1}}{\partial\lambda^{q-1}}\Bigg[\Phi\left(\lambda,\alpha,z-m+k-\frac{q}{2}+\frac{3}{2}\right)\nonumber\\
&+&\Phi\left(\lambda,\alpha,z+m-k-\frac{q}{2}+\frac{1}{2}\right)\Bigg]\Bigg|_{\lambda=1}\;.
\end{eqnarray}
and
\begin{equation}\label{10c}
f_{0,q}(\alpha,z)=\frac{2^{q-\alpha}\Gamma(\alpha)}{\Gamma(q)}\frac{\partial^{q-1}}{\partial\lambda^{q-1}}\left[\Phi\left(\lambda,\alpha,z-\frac{q}{2}+1\right)\right]\Bigg|_{\lambda=1}\;.
\end{equation}
Moreover, in the spinor field case, we will also need $f_{0,0}(\alpha,z)$ which can be easily
proved to have the expression
\begin{equation}\label{10d}
f_{0,0}(\alpha,z)=(2z)^{-\alpha}\Gamma(\alpha)\;.
\end{equation}

\section{The Effective Lagrangian for Scalar Fields}

In this section we will compute the function $\zeta_{\rm loc}(s)$ and, then, the effective Lagrangian
for scalar fields in the case of a single magnetic field, $N=1$. The final result will be obtained
in the physically relevant case of $n=4$. For convenience, let us write $\zeta_{\rm loc}(s)$
as follows
\begin{equation}\label{11}
\zeta_{\rm loc}^{\rm scalar}(s)=\zeta^{\rm scalar}_{{\rm loc}\;(0)}(s)+\zeta^{\rm scalar}_{{\rm loc}\;(1)}(s)\;,
\end{equation}
where $\zeta^{\rm scalar}_{{\rm loc}\;(0)}(s)$ and $\zeta^{\rm scalar}_{{\rm loc}\;(1)}(s)$ are, respectively, of zeroth order and first order in the Riemannian curvature.
By utilizing (\ref{4a}) with $N=1$, and by recalling the formula (\ref{5}) and the definition (\ref{10}),
we obtain, for the zeroth order part in curvature of $\zeta_{\rm loc}^{\rm scalar}(s)$,
\begin{equation}\label{12}
\zeta^{\rm scalar}_{{\rm loc}\;(0)}(s)=\frac{B^{n/2}}{(4\pi)^{n/2}\Gamma(s)}\left(\frac{M^{2}}{2z\mu^{2}}\right)^{-s}f_{0,1}\left(s-\frac{n}{2}+1,z\right)\;.
\end{equation}
Now, by using (\ref{4b}) and (\ref{4c})-(\ref{4d}) in equation (\ref{5}) we get, for the first order
part in curvature of  $\zeta_{\rm loc}^{\rm scalar}(s)$, the following expression
\begin{eqnarray}\label{13}
\zeta^{\rm scalar}_{{\rm loc}\;(1)}(s)&=&\frac{B^{n/2-1}}{(4\pi)^{n/2}\Gamma(s)}\left(\frac{M^{2}}{2z\mu^{2}}\right)^{-s}\Bigg\{\left(\frac{1}{6}R-\xi\right)f_{0,1}\left(s-\frac{n}{2}+2,z\right)\nonumber\\
&+&H^{\mu\nu}_1 R_{\mu\nu}\Bigg[\frac{1}{6}f_{0,1}\left(s-\frac{n}{2}+2,z\right)+\frac{3}{8}f_{0,1}\left(s-\frac{n}{2},z\right)-\frac{1}{24}f_{1,2}\left(s-\frac{n}{2}+3,z\right)\nonumber\\
&-&\frac{3}{8}f_{1,2}\left(s-\frac{n}{2}+1,z\right)\Bigg]+H_{1}^{\mu\nu}H_{1}^{\alpha\beta}R_{\mu\alpha\nu\beta}\Bigg[-\frac{5}{24}f_{0,1}\left(s-\frac{n}{2}+2,z\right)\nonumber\\
&-&\frac{3}{8}f_{0,1}\left(s-\frac{n}{2},z\right)+\frac{1}{24}f_{1,2}\left(s-\frac{n}{2}+3,z\right)+\frac{7}{16}f_{1,2}\left(s-\frac{n}{2}+1,z\right)\nonumber\\
&-&\frac{1}{16}f_{0,3}\left(s-\frac{n}{2}+2,z\right)\Bigg]-X_{1}^{\mu\alpha}X_{1}^{\nu\beta}R_{\mu\alpha\nu\beta}\Bigg[\frac{1}{16}f_{0,1}\left(s-\frac{n}{2}+2,z\right)\nonumber\\
&-&\frac{3}{32}f_{1,2}\left(s-\frac{n}{2}+1,z\right)+\frac{3}{32}f_{0,3}\left(s-\frac{n}{2}+2,z\right)\Bigg]\Bigg\}\;.
\end{eqnarray}

In order to obtain the one-loop effective action for scalar fields,
we need to evaluate the first derivative of $\zeta_{\rm loc}^{\rm scalar}(s)$ in (\ref{11}) at $s=0$.
It is not very difficult to realize, by inspection of the results (\ref{12}) and (\ref{13}),
that the computation of the effective Lagrangian reduces to the evaluation
of the first derivative of the function $f_{m,q}(\alpha,z)$ of specific arguments. More
precisely, we will just need to compute the following functions
\begin{equation}\label{14}
\mathcal{G}_{m,q}(n,\beta,\mu,z)=\frac{\partial}{\partial s}\left[\frac{1}{\Gamma(s)}\left(\frac{M^{2}}{2z\mu^{2}}\right)^{-s}f_{m,q}\left(s-\frac{n}{2}+\beta,z\right)\right]\Bigg|_{s=0}\;,
\end{equation}
in the three cases: $(m,q)=(0,1)$, $(m,q)=(1,2)$ and $(m,q)=(0,3)$. Here, the integer $\beta$ takes values in the set
$\{0,1,2,3\}$.

Let us begin with evaluating the derivative (\ref{14}) in the case $(m,q)=(0,1)$. By noticing that \cite{erdelyi53,gradshtein07}
\begin{equation}
\Phi(1,\alpha,z)=\zeta(\alpha,z)\;,
\end{equation}
where $\zeta(\alpha,z)$ is the Hurwitz $\zeta$-function, one obtains, in the case $n=4$, the following
expression
\begin{eqnarray}\label{15}
\mathcal{G}_{0,1}(4,\beta,\mu,z)&=&2^{3-\beta}\Bigg\{\left(\frac{M^{2}}{z\mu^{2}}\right)^{-s}\frac{\Gamma(s+\beta-2)}{\Gamma(s)}\Bigg[\Bigg(-\log\left(\frac{M^{2}}{\mu^{2}}\right)+\log\,z+\Psi(s+\beta-2)-\Psi(s)\Bigg)\nonumber\\
&\times&\zeta\left(s+\beta-2,z+\frac{1}{2}\right)+\zeta^{\prime}\left(s+\beta-2,z+\frac{1}{2}\right)\Bigg]\Bigg\}\Bigg|_{s=0}\;,
\end{eqnarray}
where $\Psi(s)$ represents the digamma function and $\zeta^{\prime}(s,\alpha)=\frac{d}{ds}\zeta(s,\alpha)$.
From equations (\ref{12}) and (\ref{13}), one can easily see that we need the expression (\ref{15}) for the following particular values of the parameter $\beta$:

For $\beta=0$, one obtains
\begin{equation}
\mathcal{G}_{0,1}(4,0,\mu,z)=\frac{z}{3}(1-4z^{2})\left[-\log\left(\frac{M^{2}}{\mu^{2}}\right)+\log\,z+\frac{3}{2}\right]+4\zeta^{\prime}\left(-2,z+\frac{1}{2}\right)\;.
\end{equation}

For $\beta=1$, one obtains
\begin{equation}
\mathcal{G}_{0,1}(4,1,\mu,z)=\frac{1}{6}(1-12z^{2})\left[\log\left(\frac{M^{2}}{\mu^{2}}\right)-\log\,z-1\right]-4\zeta^{\prime}\left(-1,z+\frac{1}{2}\right)\;.
\end{equation}

For $\beta=2$, one obtains
\begin{equation}
\mathcal{G}_{0,1}(4,2,\mu,z)=2z\left[\log\left(\frac{M^{2}}{\mu^{2}}\right)-\log\,z\right]+2\zeta^{\prime}\left(0,z+\frac{1}{2}\right)\;.
\end{equation}

In order to arrive at the formulas above we have used the well known result \cite{erdelyi53,gradshtein07} stating that for any integer $u\geq 0$
\begin{equation}\label{15a}
\zeta(-u,p)=-\frac{B_{u+1}(p)}{u+1}\;,
\end{equation}
where $B_{u}(p)$ are the Bernoulli polynomials.

Let us evaluate, now, the derivative (\ref{14}) in the case $(m,q)=(1,2)$. It is not difficult to
show, by differentiating the series in (\ref{8a}), that
\begin{equation}\label{16}
\left\{\frac{\partial}{\partial x}\Phi(x,\gamma,\nu)\right\}\Bigg|_{x=1}=\zeta(\gamma-1,\nu)-\nu\zeta(\gamma,\nu)\;.
\end{equation}
By utilizing the relation (\ref{16}), we have, in the case $n=4$, the expression
\begin{eqnarray}\label{17}
\mathcal{G}_{1,2}(4,\beta,\mu,z)&=&2^{3-\beta}\Bigg\{\left(\frac{M^{2}}{z\mu^{2}}\right)^{-s}\frac{\Gamma(s+\beta-2)}{\Gamma(s)}\Bigg[\Bigg(-\log\left(\frac{M^{2}}{\mu^{2}}\right)+\log\,z+\Psi(s+\beta-2)-\Psi(s)\Bigg)\nonumber\\
&\times&\Omega(s,\beta,z)+\Omega^{\prime}(s,\beta,z)\Bigg]\Bigg\}\Bigg|_{s=0}\;,
\end{eqnarray}
where $\Omega^{\prime}(s,\beta,z)=\frac{d}{ds}\Omega(s,\beta,z)$, and $\Omega(s,\beta,z)$ is an auxiliary function
defined as
\begin{eqnarray}
\Omega(s,\beta,z)&=&\zeta\left(s+\beta-3,z-\frac{1}{2}\right)-\left(z-\frac{1}{2}\right)\zeta\left(s+\beta-2,z-\frac{1}{2}\right)\nonumber\\
&+&\zeta\left(s+\beta-3,z+\frac{1}{2}\right)-\left(z+\frac{1}{2}\right)\zeta\left(s+\beta-2,z+\frac{1}{2}\right)\;.
\end{eqnarray}
By inspection of (\ref{12}) and (\ref{13}), one easily realizes that we will need to compute equation (\ref{17}) for the following
values of the parameter $\beta$:


For $\beta=1$ one gets
\begin{eqnarray}\label{18}
\mathcal{G}_{1,2}(4,1,\mu,z)&=&\frac{z}{3}(1+4z^{2})\left[\log\left(\frac{M^{2}}{\mu^{2}}\right)-\log\,z-1\right]-4\Bigg[\zeta^{\prime}\left(-2,z-\frac{1}{2}\right)+\zeta^{\prime}\left(-2,z+\frac{1}{2}\right)\nonumber\\
&-&\left(z-\frac{1}{2}\right)\zeta^{\prime}\left(-1,z-\frac{1}{2}\right)-\left(z+\frac{1}{2}\right)\zeta^{\prime}\left(-1,z+\frac{1}{2}\right)\Bigg]\;.
\end{eqnarray}

For $\beta=3$ one gets
\begin{eqnarray}\label{19}
\mathcal{G}_{1,2}(4,3,\mu,z)&=&1+2z\left[\log\left(\frac{M^{2}}{\mu^{2}}\right)-\log\,z-1\right]+z\left[\Psi\left(z-\frac{1}{2}\right)+\Psi\left(z+\frac{1}{2}\right)\right]\nonumber\\
&-&\frac{1}{2}\left[\Psi\left(z-\frac{1}{2}\right)-\Psi\left(z+\frac{1}{2}\right)\right]\;.
\end{eqnarray}

Here, we have used the relation (\ref{15a}) to obtain (\ref{18}) and (\ref{19}).

The only case left to consider, for scalar fields, is $(m,q)=(0,3)$. By noticing that the following relation holds
\begin{equation}
\left\{\frac{\partial^{2}}{\partial x^{2}}\Phi(x,\gamma,\nu)\right\}\Bigg|_{x=1}=\zeta(\gamma-2,\nu)-(2\nu+1)\zeta(\gamma-1,\nu)+\nu(\nu+1)\zeta(\gamma,\nu)\;,
\end{equation}
we have, in the case $n=4$, the expression
\begin{eqnarray}\label{20}
\mathcal{G}_{0,3}(4,\beta,\mu,z)&=&2^{4-\beta}\Bigg\{\left(\frac{M^{2}}{z\mu^{2}}\right)^{-s}\frac{\Gamma(s+\beta-2)}{\Gamma(s)}\Bigg[\Bigg(-\log\left(\frac{M^{2}}{\mu^{2}}\right)+\log\,z+\Psi(s+\beta-2)-\Psi(s)\Bigg)\nonumber\\
&\times&\Theta(s,\beta,z)+\Theta^{\prime}(s,\beta,z)\Bigg]\Bigg\}\Bigg|_{s=0}\;,
\end{eqnarray}
where where $\Theta^{\prime}(s,\beta,z)=\frac{d}{ds}\Theta(s,\beta,z)$, and $\Theta(s,\beta,z)$ is defined as
\begin{eqnarray}
\Theta(s,\beta,z)&=&\zeta\left(s+\beta-4,z-\frac{1}{2}\right)-2z\,\zeta\left(s+\beta-3,z-\frac{1}{2}\right)+\left(z^{2}-\frac{1}{4}\right)\zeta\left(s+\beta-2,z-\frac{1}{2}\right)\,.\;\;\;\;\;\;\;\;
\end{eqnarray}
By recalling (\ref{12}) and (\ref{13}) we soon realize that we need to compute (\ref{20}) for only one value of
the parameter $\beta$, namely:

For $\beta=2$, by using (\ref{15a}), we obtain
\begin{eqnarray}
\mathcal{G}_{0,3}(4,2,\mu,z)&=&-\frac{z}{3}(3-4z^{2})\left[\log\left(\frac{M^{2}}{\mu^{2}}\right)-\log\,z\right]+4\Bigg[\zeta^{\prime}\left(-2,z-\frac{1}{2}\right)\nonumber\\
&-&2z\,\zeta^{\prime}\left(-1,z-\frac{1}{2}\right)+\left(z^{2}-\frac{1}{4}\right)\zeta^{\prime}\left(0,z-\frac{1}{2}\right)\Bigg]\;.
\end{eqnarray}


By using the explicit formulas for $\mathcal{G}_{m,q}(n,\beta,\mu,z)$ found in this section, we are finally
able to write the (unrenormalized) effective Lagrangian for scalar fields in $n=4$. Explicitly, we have
\begin{eqnarray}\label{200}
\mathcal{L}^{\,\rm scalar}&=&-\frac{B^{2}}{16\pi^{2}}h_{1}\left(\frac{M^{2}}{\mu^{2}},z\right)-\frac{B}{16\pi^{2}}\Bigg[\left(\frac{1}{6}-\xi\right)R\,h_{2}\left(\frac{M^{2}}{\mu^{2}},z\right)+H_{1}^{\mu\nu}R_{\mu\nu}\,h_{3}\left(z\right)\nonumber\\
&+&H_{1}^{\mu\nu}H_{1}^{\alpha\beta}R_{\mu\alpha\nu\beta}\,h_{4}\left(z\right)-X_{1}^{\mu\alpha}X_{1}^{\nu\beta}R_{\mu\alpha\nu\beta}\,h_{5}\left(z\right)\Bigg]\;,
\end{eqnarray}
where the functions $h_{i}$ are found to have the following expressions
\begin{eqnarray}
h_{1}\left(\frac{M^{2}}{\mu^{2}},z\right)&=&\left(\frac{1}{6}-2z^{2}\right)\left[\log\left(\frac{M^{2}}{\mu^{2}}\right)-\log\,z-1\right]-4\zeta^{\prime}\left(-1,z+\frac{1}{2}\right)\;,\\[10pt]
h_{2}\left(\frac{M^{2}}{\mu^{2}},z\right)&=&2z\left[\log\left(\frac{M^{2}}{\mu^{2}}\right)-\log\,z\right]+2\zeta^{\prime}\left(0,z+\frac{1}{2}\right)\;,\\[10pt]
h_{3}\left(z\right)&=&-\frac{1}{24}+\frac{z}{16}\left(\frac{19}{3}-4z^{2}\right)-\frac{z}{24}\left[\Psi\left(z-\frac{1}{2}\right)+\Psi\left(z+\frac{1}{2}\right)\right]\nonumber\\
&+&\frac{1}{48}\left[\Psi\left(z-\frac{1}{2}\right)-\Psi\left(z+\frac{1}{2}\right)\right]+\frac{1}{3}\zeta^{\prime}\left(0,z+\frac{1}{2}\right)+3\zeta^{\prime}\left(-2,z+\frac{1}{2}\right)\nonumber\\
&+&\frac{3}{2}\zeta^{\prime}\left(-2,z-\frac{1}{2}\right)-\frac{3}{2}\left(z-\frac{1}{2}\right)\zeta^{\prime}\left(-1,z-\frac{1}{2}\right)-\frac{3}{2}\left(z+\frac{1}{2}\right)\zeta^{\prime}\left(-1,z+\frac{1}{2}\right)\;,
\end{eqnarray}
\begin{eqnarray}
h_{4}\left(z\right)&=&\frac{1}{24}-\frac{z}{12}\Big(5-2z^{2}\Big)+\frac{z}{24}\left[\Psi\left(z-\frac{1}{2}\right)+\Psi\left(z+\frac{1}{2}\right)\right]\nonumber\\
&-&\frac{1}{48}\left[\Psi\left(z-\frac{1}{2}\right)-\Psi\left(z+\frac{1}{2}\right)\right]-\frac{5}{12}\zeta^{\prime}\left(0,z+\frac{1}{2}\right)-\frac{1}{4}\left(z^{2}-\frac{1}{4}\right)\zeta^{\prime}\left(0,z-\frac{1}{2}\right)\nonumber\\
&-&\frac{13}{4}\zeta^{\prime}\left(-2,z+\frac{1}{2}\right)-2\zeta^{\prime}\left(-2,z-\frac{1}{2}\right)+\frac{1}{4}\left(9z-\frac{7}{2}\right)\zeta^{\prime}\left(-1,z-\frac{1}{2}\right)\nonumber\\
&+&\frac{7}{4}\left(z+\frac{1}{2}\right)\zeta^{\prime}\left(-1,z+\frac{1}{2}\right)\;,\\[10pt]
h_{5}\left(z\right)&=&\frac{z}{32}\Big(1+4z^{2}\Big)+\frac{1}{8}\zeta^{\prime}\left(0,z+\frac{1}{2}\right)+\frac{3}{8}\left(z^{2}-\frac{1}{4}\right)\zeta^{\prime}\left(0,z-\frac{1}{2}\right)\nonumber\\
&+&\frac{3}{4}\zeta^{\prime}\left(-2,z-\frac{1}{2}\right)+\frac{3}{8}\zeta^{\prime}\left(-2,z+\frac{1}{2}\right)-\frac{3}{8}\left(3z-\frac{1}{2}\right)\zeta^{\prime}\left(-1,z-\frac{1}{2}\right)\nonumber\\
&-&\frac{3}{8}\left(z+\frac{1}{2}\right)\zeta^{\prime}\left(-1,z+\frac{1}{2}\right)\;.
\end{eqnarray}
As we can see, from these formulas, $h_{3}$, $h_{4}$ and $h_{5}$ do not depend on the renormalization parameter $\mu$
as one should expect. The functions $h_{1}$ and $h_{2}$, instead, contain $\mu$ and they can be used in order to
renormalize the coupling constants in the Einstein-Maxwell action. The function $h_{1}$, found above, represents the
standard result for the effective Lagrangian for scalar fields obtained in a spacetime with vanishing curvature \cite{dunne04}.

\section{The Effective Lagrangian for Spinor Fields}

In this section we turn our attention to the spinor field case. With arguments similar to the ones
used in the previous section, we will present the function $\zeta_{\rm loc}(s)$ and, subsequently, the
effective Lagrangian for spinors. We will assume, as before, that only one magnetic field is present, $N=1$.
As we have already done in (\ref{11}), we split $\zeta_{\rm loc}^{\rm spinor}(s)$ as follows
\begin{equation}
\zeta_{\rm loc}^{\rm spinor}(s)=\zeta^{\rm spinor}_{{\rm loc}\;(0)}(s)+\zeta^{\rm spinor}_{{\rm loc}\;(1)}(s)\;.
\end{equation}
By recalling the equations (\ref{4f}), (\ref{5}) and by using the result (\ref{10b}) we obtain
\begin{equation}\label{20a}
\zeta^{\rm spinor}_{{\rm loc}\;(0)}(s)=\frac{2^{[n/2]}B^{n/2}}{(4\pi)^{n/2}\Gamma(s)}\left(\frac{M^{2}}{2z\mu^{2}}\right)^{-s}f_{1,1}\left(s-\frac{n}{2}+1,z\right)\;.
\end{equation}
At this point, by substituting (\ref{4c}), (\ref{4d}), (\ref{4g}) in the expression (\ref{5})
and by recalling the results (\ref{10a})-(\ref{10c}), we have for $\zeta^{\rm spinor}_{{\rm loc}\;(1)}(s)$ the formula
\begin{eqnarray}\label{20b}
\zeta^{\rm spinor}_{{\rm loc}\;(1)}(s)&=&\frac{2^{[n/2]}B^{n/2-1}}{(4\pi)^{n/2}\Gamma(s)}\left(\frac{M^{2}}{2z\mu^{2}}\right)^{-s}\Bigg\{-\frac{1}{12}Rf_{1,1}\left(s-\frac{n}{2}+2,z\right)\nonumber\\
&+&H_{1}^{\mu\nu}R_{\mu\nu}\Bigg[\frac{1}{6}f_{1,1}\left(s-\frac{n}{2}+2,z\right)+\frac{3}{8}f_{1,1}\left(s-\frac{n}{2},z\right)-\frac{1}{24}f_{2,2}\left(s-\frac{n}{2}+3,z\right)\nonumber\\
&-&\frac{3}{8}f_{2,2}\left(s-\frac{n}{2}+1,z\right)\Bigg]+H_{1}^{\mu\nu}H_{1}^{\alpha\beta}R_{\mu\alpha\nu\beta}\Bigg[-\frac{5}{24}f_{1,1}\left(s-\frac{n}{2}+2,z\right)\nonumber\\
&-&\frac{3}{8}f_{1,1}\left(s-\frac{n}{2},z\right)+\frac{1}{24}f_{2,2}\left(s-\frac{n}{2}+3,z\right)+\frac{7}{16}f_{2,2}\left(s-\frac{n}{2}+1,z\right)\nonumber\\
&-&\frac{1}{16}f_{1,3}\left(s-\frac{n}{2}+2,z\right)\Bigg]-X_{1}^{\mu\alpha}X_{1}^{\nu\beta}R_{\mu\alpha\nu\beta}\Bigg[-\frac{3}{16}f_{1,1}\left(s-\frac{n}{2}+2,z\right)\nonumber\\
&-&\frac{3}{32}f_{2,2}\left(s-\frac{n}{2}+1,z\right)+\frac{3}{32}f_{1,3}\left(s-\frac{n}{2}+2,z\right)+\frac{1}{4}f_{0,0}\left(s-\frac{n}{2}+1,z\right)\Bigg]\Bigg\}\;.
\end{eqnarray}
In order to evaluate the effective action, we need to compute the derivatives (\ref{14}) in the following four cases:
$(m,q)=(1,1)$, $(m,q)=(2,2)$, $(m,q)=(1,3)$ and lastly $(m,q)=(0,0)$.

By following the same arguments used in the previous section, we find, for the case $(m,q)=(1,1)$ in $n=4$,
\begin{eqnarray}\label{21}
\mathcal{G}_{1,1}(4,\beta,\mu,z)&=&2^{2-\beta}\Bigg\{\left(\frac{M^{2}}{z\mu^{2}}\right)^{-s}\frac{\Gamma(s+\beta-2)}{\Gamma(s)}\Bigg[\Bigg(-\log\left(\frac{M^{2}}{\mu^{2}}\right)+\log\,z+\Psi(s+\beta-2)-\Psi(s)\Bigg)\nonumber\\
&\times&\Big[2\zeta\left(s+\beta-2,z\right)-z^{-s-\beta+2}\Big]+2\zeta^{\prime}\left(s+\beta-2,z\right)+z^{-s-\beta+2}\log\,z\Bigg]\Bigg\}\Bigg|_{s=0}\;,
\end{eqnarray}
where we have used the relation, valid for any integer $p\geq 1$,
\begin{equation}\label{22}
\zeta(s,v+p)=\zeta(s,v)-\sum_{n=0}^{p-1}\frac{1}{(v+n)^{s}}\;,
\end{equation}
to obtain the expression (\ref{21}). From (\ref{20a}) and (\ref{20b}) one can easily see that we will need
the expression (\ref{21}) evaluated for the following particular values of $\beta$:


For $\beta=0$, one has
\begin{equation}
\mathcal{G}_{1,1}(4,0,\mu,z)=\frac{2}{3}z(1+2z^{2})\left[\log\left(\frac{M^{2}}{\mu^{2}}\right)-\log\,z-\frac{3}{2}\right]+4\zeta^{\prime}(-2,z)+2z^{2}\log\,z\;.
\end{equation}

For $\beta=1$, one has
\begin{equation}
\mathcal{G}_{1,1}(4,1,\mu,z)=\frac{1}{3}(1+6z^{2})\left[-\log\left(\frac{M^{2}}{\mu^{2}}\right)+\log\,z+1\right]-4\zeta^{\prime}(-1,z)-2z\log\,z\;.
\end{equation}

For $\beta=2$, one has
\begin{equation}
\mathcal{G}_{1,1}(4,2,\mu,z)=2z\left[\log\left(\frac{M^{2}}{\mu^{2}}\right)-\log\,z\right]+2\zeta^{\prime}(0,z)+\log\,z\;.
\end{equation}

To obtain the previous equations we have used the relation (\ref{15a}).

We turn, at this point, our attention to the next relevant case, namely $(m,q)=(2,2)$. It is straightforward
to show that, in $n=4$,
\begin{eqnarray}
\mathcal{G}_{2,2}(4,\beta,\mu,z)&=&2^{2-\beta}\Bigg\{\left(\frac{M^{2}}{z\mu^{2}}\right)^{-s}\frac{\Gamma(s+\beta-2)}{\Gamma(s)}\Bigg[\Bigg(-\log\left(\frac{M^{2}}{\mu^{2}}\right)+\log\,z+\Psi(s+\beta-2)-\Psi(s)\Bigg)\nonumber\\[10pt]
&\times&\Big[4\zeta(s+\beta-3,z)-4z\,\zeta(s+\beta-2,z)+z^{-s-\beta+2}\Big]+4\zeta^{\prime}(s+\beta-3,z)\nonumber\\[10pt]
&-&4z\,\zeta^{\prime}(s+\beta-2,z)-z^{-s-\beta+2}\log\,z\Bigg]\Bigg\}\Bigg|_{s=0}\;,
\end{eqnarray}
where we have used the relation (\ref{22}) and the result
\begin{equation}
\zeta(s,v-p)=\zeta(s,v)+\sum_{n=0}^{p-1}\frac{1}{(v-p+n)^{s}}\;,
\end{equation}
valid for any integer $p\geq 0$. From (\ref{20a}) and (\ref{20b}), we need the following particular values of
the parameter $\beta$:


For $\beta=1$, we have
\begin{eqnarray}
\mathcal{G}_{2,2}(4,1,\mu,z)&=&\frac{4}{3}z(1+z^{2})\left[\log\left(\frac{M^{2}}{\mu^{2}}\right)-\log\,z-1\right]-8\zeta^{\prime}(-2,z)\nonumber\\[10pt]
&+&8z\,\zeta^{\prime}(-1,z)+2z\log\,z\;.
\end{eqnarray}

For $\beta=3$, we have
\begin{equation}
\mathcal{G}_{2,2}(4,3,\mu,z)=1+2z\left[\log\left(\frac{M^{2}}{\mu^{2}}\right)-\log\,z\right]+\frac{1-4z^{2}}{2z}+2z\,\Psi(z)\;.
\end{equation}


The next case that we need to consider for spinor fields is $(m,q)=(1,3)$. In $n=4$, one can show that
\begin{eqnarray}
\mathcal{G}_{1,3}(4,\beta,\mu,z)&=&2^{3-\beta}\Bigg\{\left(\frac{M^{2}}{z\mu^{2}}\right)^{-s}\frac{\Gamma(s+\beta-2)}{\Gamma(s)}\Bigg[\Bigg(-\log\left(\frac{M^{2}}{\mu^{2}}\right)+\log\,z+\Psi(s+\beta-2)-\Psi(s)\Bigg)\nonumber\\[10pt]
&\times&\Big[2\zeta(s+\beta-4,z)-4z\,\zeta(s+\beta-3,z)+2z^{2}\zeta(s+\beta-2,z)\Big]+2\zeta^{\prime}(s+\beta-4,z)\nonumber\\[10pt]
&-&4z\,\zeta^{\prime}(s+\beta-3,z)+2z^{2}\zeta^{\prime}(s+\beta-2,z)\Bigg]\Bigg\}\Bigg|_{s=0}\;,
\end{eqnarray}
Now, we only need to evaluate $\mathcal{G}_{1,3}(4,\beta,\mu,z)$ for only one value of $\beta$. More specifically, by using (\ref{15a}) we obtain:


For $\beta=2$,
\begin{equation}
\mathcal{G}_{1,3}(4,2,\mu,z)=\frac{4}{3}z^{3}\left[\log\left(\frac{M^{2}}{\mu^{2}}\right)-\log\,z\right]+4\zeta^{\prime}(-2,z)-8z\,\zeta^{\prime}(-1,z)+4z^{2}\zeta^{\prime}(0,z)\;.
\end{equation}


The last derivative that we need to complete the spinor field case is $\mathcal{G}_{0,0}(4,\beta,\mu,z)$ solely for $\beta=1$.
From (\ref{10d}) and the definition (\ref{14}), it is straightforward to show that
\begin{equation}
\mathcal{G}_{0,0}(4,1,\mu,z)=2z\left[\log\left(\frac{M^{2}}{\mu^{2}}\right)-1\right]\;.
\end{equation}

We finally have all the formulas that we need in order to write the unrenormalized
effective Lagrangian for spinor fields in $n=4$. By utilizing the derivatives $\mathcal{G}_{m,q}(4,\beta,\mu,z)$
found in this section we obtain
\begin{eqnarray}\label{23}
\mathcal{L}^{\,\rm spinor}&=&\frac{B^{2}}{8\pi^{2}}w_{1}\left(\frac{M^{2}}{\mu^{2}},z\right)+\frac{B}{8\pi^{2}}\Bigg[-\frac{1}{12}R\,w_{2}\left(\frac{M^{2}}{\mu^{2}},z\right)+H_{1}^{\mu\nu}R_{\mu\nu}\,w_{3}\left(z\right)\nonumber\\
&+&H_{1}^{\mu\nu}H_{1}^{\alpha\beta}R_{\mu\alpha\nu\beta}\,w_{4}\left(z\right)-X_{1}^{\mu\alpha}X_{1}^{\nu\beta}R_{\mu\alpha\nu\beta}\,w_{5}\left(z\right)\Bigg]\;,
\end{eqnarray}
where the functions $w_{i}$ are defined as follows
\begin{eqnarray}\label{24}
w_{1}\left(\frac{M^{2}}{\mu^{2}},z\right)&=&\frac{1}{3}(1+6z^{2})\left[-\log\left(\frac{M^{2}}{\mu^{2}}\right)+\log\,z+1\right]-4\zeta^{\prime}(-1,z)-2z\log\,z\;,\\[10pt]
w_{2}\left(\frac{M^{2}}{\mu^{2}},z\right)&=&2z\left[\log\left(\frac{M^{2}}{\mu^{2}}\right)-\log\,z\right]+2\zeta^{\prime}(0,z)+\log\,z\;,\\[10pt]
w_{3}\left(z\right)&=&-\frac{1}{24}+\frac{1}{6}\log\,z+\frac{3}{4}z(z-1)\log\,z-\frac{1}{48z}+\frac{z}{24}(5-6z^{2})-\frac{z}{12}\Psi(z)\nonumber\\
&+&\frac{9}{2}\zeta^{\prime}(-2,z)-3z\,\zeta^{\prime}(-1,z)+\frac{1}{3}\zeta^{\prime}(0,z)\;,
\end{eqnarray}
\begin{eqnarray}
w_{4}\left(z\right)&=&\frac{1}{24}-\frac{5}{24}\log\,z+\frac{z}{8}(7-6z)\log\,z+\frac{1}{48z}-\frac{z}{24}(7-4z^{2})+\frac{z}{12}\Psi(z)\nonumber\\
&-&\frac{21}{4}\zeta^{\prime}(-2,z)+4z\,\zeta^{\prime}(-1,z)-\frac{1}{12}(5+3z^{2})\zeta^{\prime}(0,z)\;,\\[10pt]
w_{5}\left(z\right)&=&-\frac{1}{16}(3-5z)\log\,z-\frac{z}{8}(3-z^{2})+\frac{9}{8}\zeta^{\prime}(-2,z)-\frac{3}{2}z\,\zeta^{\prime}(-1,z)\nonumber\\
&-&\frac{3}{8}(1-z^{2})\zeta^{\prime}(0,z)\;.
\end{eqnarray}
Also in this case, the functions $w_{3}$, $w_{4}$ and $w_{5}$ do not depend on the
renormalization parameter $\mu$ as one should expect. The result for $w_{1}$ represents the well known
unrenormalized effective Lagrangian for spinor fields in flat spacetime under the influence of solely
a covariantly constant magnetic field (see e.g. \cite{dunne04}).

\section{Massless Scalar and Spinor Fields}

In this section we will study the effective Lagrangian in the limit of massless
scalar and spinor fields. As one can understand from the results obtained in the previous
sections, this limit is realized when the variable $z\to 0$, which means that we are
actually considering the regime
\begin{equation}
M^{2}\ll B\;.
\end{equation}
It is obvious, from the last formula, that the massless limit is equivalent to
the case in which a very strong background magnetic field is present.

Before presenting the results for the effective Lagrangian in the massless case,
we will compute the limit as $z\to 0$ of the Hurwitz $\zeta$-functions appearing in the
functions $h_{i}$ and $w_{i}$.

Let us start with the evaluation of the quantity $\zeta^{\prime}(s,1/2)$. It is not difficult to show,
by using the multiplication theorem for the Hurwitz $\zeta$-function,
that
\begin{equation}\label{24a}
\zeta\left(s,\frac{1}{2}\right)=(2^{2}-1)\zeta(s)\;.
\end{equation}
By differentiating the above equation we obtain the desired formula for the derivative
\begin{equation}\label{25}
\zeta^{\prime}\left(s,\frac{1}{2}\right)=2^{s}\zeta(s)\log\,2+(2^{2}-1)\zeta^{\prime}(s)\;,
\end{equation}
where $\zeta(s)$ is the Riemann $\zeta$-function and $\zeta^{\prime}(s)=\frac{d}{ds}\zeta(s)$.
The first derivative of the Riemann $\zeta$-function appearing in (\ref{25}) can be
computed by utilizing the functional relation \cite{gradshtein07,erdelyi53}
\begin{equation}\label{26}
\zeta(1-s)=2(2\pi)^{-s}\Gamma(s)\cos\left(\frac{\pi s}{2}\right)\zeta(s)\;.
\end{equation}
By differentiating (\ref{26}) with respect to the variable $s$ and then setting $s=2k+1$, for $k\geq 1$,
one obtains
\begin{equation}
\zeta^{\prime}(-2k)=(-1)^{k}\pi(2\pi)^{-2k-1}\Gamma(2k+1)\zeta(2k+1)\;.
\end{equation}
In particular we will need the following value
\begin{equation}\label{27}
\zeta^{\prime}(-2)=-\frac{1}{4\pi^{2}}\zeta(3)\;.
\end{equation}

Now, by differentiating (\ref{26}) with respect to $s$ and by setting $s=2k$, for $k\geq 1$,
one has an expression for the derivative of $\zeta(s)$ at negative integer points \cite{miller98}
\begin{equation}\label{28}
\zeta^{\prime}(-2k+1)=\frac{B_{2k}}{2k}\Big(\Psi(2k)-\log\,2\pi\Big)+\frac{(-1)^{k+1}2(2k-1)!}{(2\pi)^{2k}}\zeta^{\prime}(2k)\;,
\end{equation}
where $B_{2k}$ represent the Bernoulli numbers.
In particular, we will use the equation (\ref{28}) for the particular value of $k=1$
\begin{equation}\label{29}
\zeta^{\prime}(-1)=\frac{1}{12}-\log\,A\;,
\end{equation}
where $A$ is the Glaisher-Kinkelin constant \cite{finch03}. We would like to point out that in order to obtain the result
(\ref{29}) we have used the relation
\begin{equation}
\zeta^{\prime}(2)=\frac{\pi^{2}}{6}\log\,2\pi+\frac{\pi^{2}}{6}\gamma-2\pi^{2}\log\,A\;,
\end{equation}
where $\gamma$ is the Euler-Mascheroni constant, which can be straightforwardly proved by using the definition \cite{finch03}
\begin{equation}
A=(2\pi)^{\frac{1}{12}}\Bigg[\exp\left\{\frac{\gamma \pi^{2}}{6}-\zeta^{\prime}(2)\right\}\Bigg]^{\frac{1}{2\pi^{2}}}\;.
\end{equation}

For the purpose of this paper we will need the following particular values
\begin{eqnarray}\label{210}
\zeta^{\prime}\left(0,\frac{1}{2}\right)&=&-\frac{1}{2}\log\,2\;,\\
\zeta^{\prime}\left(-1,\frac{1}{2}\right)&=&-\frac{1}{24}-\frac{1}{24}\log\,2+\frac{1}{2}\log\,A\;,\\
\zeta^{\prime}\left(-2,\frac{1}{2}\right)&=&\frac{3}{16\pi^{2}}\zeta(3)\;,\label{211}
\end{eqnarray}
where we have used the relations (\ref{15a}), (\ref{27}), (\ref{29}) and the fact that \cite{gradshtein07,erdelyi53} $\zeta^{\prime}(0)=-1/2\log\,2\pi$.

For the scalar field case, we will also need to compute the quantity $\zeta^{\prime}(s,-1/2)$. One can prove,
by utilizing the defining series of the Hurwitz $\zeta$-function, that
\begin{equation}\label{29a}
\zeta\left(s,q-\frac{1}{2}\right)=\frac{2^{s}}{(2q-1)^{s}}+2^{s}\zeta(s,2q)-\zeta(s,q)\;.
\end{equation}
It follows from (\ref{29a}) that, for vanishing $q$,
\begin{equation}\label{30}
\zeta\left(s,-\frac{1}{2}\right)=(-2)^{s}+(2^{s}-1)\zeta(s)\;.
\end{equation}
By differentiating (\ref{30}) we get
\begin{equation}\label{31}
\zeta^{\prime}\left(s,-\frac{1}{2}\right)=2^{s}\Big[(-1)^{s}\Big(i\pi+\log\,2\Big)+\zeta(s)\log\,2\Big]+(2^{s}-1)\zeta^{\prime}(s)\;.
\end{equation}

In particular, to analyze the massless limit, we will employ the following specific
values of (\ref{31})
\begin{eqnarray}\label{31a}
\zeta^{\prime}\left(0,-\frac{1}{2}\right)&=&i\pi+\frac{1}{2}\log\,2\;,\\
\zeta^{\prime}\left(-1,-\frac{1}{2}\right)&=&-\frac{i\pi}{2}-\frac{1}{24}-\frac{13}{24}\log\,2+\frac{1}{2}\log\,A\;,\\
\zeta^{\prime}\left(-2,-\frac{1}{2}\right)&=&\frac{i\pi}{4}+\frac{1}{4}\log\,2+\frac{3}{16\pi^{2}}\zeta(3)\;,\label{31b}
\end{eqnarray}
where we have exploited the results (\ref{15a}), (\ref{27}) and (\ref{29}).

Let us, at this point, focus our attention to the spinor case. In order to analyze the massless
limit we need to evaluate $\zeta^{\prime}(s,z)$ when $z\to 0$ for particular values of $s$.
By taking the first derivative of the defining series of the Hurwitz $\zeta$-function we obtain
\begin{equation}\label{32}
\zeta^{\prime}(s,z)=-z^{-s}\log\,z-\sum_{n=1}^{\infty}(n+z)^{-s}\log\,(n+z)\;.
\end{equation}
The limit as $z\to 0$ of the above quantity is finite for any $s\leq -1$ and leads to the expression
\begin{equation}\label{32a}
\zeta^{\prime}(-s,0)=\zeta^{\prime}(-s)\;.
\end{equation}
For $s=0$, instead, the derivative (\ref{32}) takes the following form
\begin{equation}\label{33}
\zeta^{\prime}(0,z)=-\log\,z+\zeta^{\prime}(0)\;,
\end{equation}
which presents a logarithmic divergence for $z$ approaching zero.

Now that we have determined the limit as $z\to 0$ of the various Hurwitz $\zeta$-functions
which appear in the results for the effective Lagrangian for scalar and spinor fields,
we can proceed with the analysis of the massless case.

\subsection{Scalar Fields}

The effective Lagrangian for massless scalar fields in $n=4$ has the form (\ref{200})
where, in the limit $z\to 0$, the functions $h_{i}$ are found to be
\begin{eqnarray}
h_{1}\left(\frac{B}{\mu^{2}}\right)&=&\frac{1}{6}\log\left(\frac{B}{\mu^{2}}\right)+\frac{1}{3}\log\,2-2\log\,A\;,\\[10pt]
h_{2}&=&-\log\,2\;,\\[10pt]
h_{3}&=&-\frac{1}{6}\log\,2+\frac{27}{32\pi^{2}}\zeta(3)\;,\\[10pt]
h_{4}&=&\frac{17}{96}\log\,2-\frac{63}{64\pi^{2}}\zeta(3)\;,\\[10pt]
h_{5}&=&-\frac{1}{64}\log\,2+\frac{27}{128\pi^{2}}\zeta(3)\;.
\end{eqnarray}
In order to obtain the last equations we have used the results (\ref{210})-(\ref{211}) and (\ref{31a})-(\ref{31b}).
Notice that although the derivatives (\ref{31a})-(\ref{31b}) are imaginary quantities, the particular form
of the functions $h_{i}$ leads correctly to a result which is real.

\subsection{Spinor Fields}

For massless spinor fields the effective Lagrangian has the form (\ref{23}). However,
in the limit of vanishing mass the functions $w_{i}$ become
\begin{eqnarray}
w_{1}\left(\frac{B}{\mu^{2}}\right)&=&-\frac{1}{3}\log\left(\frac{B}{\mu^{2}}\right)-\frac{1}{3}\log\,2+4\log\,A\;,\\[10pt]
w_{2}(z)&=&-\log\,z-\frac{1}{2}\log\,2\pi\;,\\[10pt]
w_{3}(z)&=&-\frac{1}{48z}-\frac{1}{6}\log\,z+\frac{5}{72}-\frac{9}{8\pi^{2}}\zeta(3)-\frac{1}{3}\log\,A\;,\\[10pt]
w_{4}(z)&=&\frac{1}{48z}+\frac{5}{24}\log\,z+\frac{1}{144}+\frac{21}{16\pi^{2}}\zeta(3)+\frac{5}{12}\log\,A\;,\\[10pt]
w_{5}(z)&=&\frac{3}{16}\log\,z-\frac{1}{32}-\frac{9}{32\pi^{2}}\zeta(3)+\frac{3}{8}\log\,A\;,
\end{eqnarray}
where we have exploited the results (\ref{32a}) and (\ref{33}).

As we can see from this last result
the massless limit of the term of zeroth order in curvature $w_{1}$ is finite. The terms contributing
to the linear part in curvature of the effective Lagrangian clearly contain infrared divergences
of the type $M^{-2}$ and $\log\,M^{2}$. This particular behavior is due to the fact that in $n=4$
the function $tB \coth(tB)$, which contributes to $\mathcal{B}_{1}^{\rm spinor}$,
does not provide a cut-off for the integral (\ref{5a}) for large $t$ in absence of mass.
We would also like to mention, here, that the presence of the infrared divergences depends on
the dimensionality of the spacetime. In fact a quick analysis of the integrals leading to the
infrared divergent terms shows that for $n\geq 4$ the massless limit for the linear part
in curvature of the effective Lagrangian for spinor fields is finite.

\section{Concluding Remarks}

In this paper we have investigated the effective action for complex scalar and spinor fields
under the influence of a strong background magnetic field in curved spacetime. We have analyzed, here,
an essentially \emph{non-perturbative} regime in which the covariantly constant background magnetic field
is so strong that all its orders have to be taken into account. In order to
compute the one-loop effective action for scalar and spinor fields in this setting
we have used the method of $\zeta$-function regularization. The spectral $\zeta$-function for the Laplace type operator
under investigation has been represented in terms of the heat kernel, for which
we have employed a non-perturbative asymptotic expansion recently found in \cite{avramidi08e}.

Here, we have explicitly found the gravitational corrections, linear in the Riemannian
curvature and to \emph{all orders in the magnetic field}, to the effective Lagrangian for
scalar and spinor fields in the physically relevant case of a four dimensional spacetime.
To the best of our knowledge, this represents a completely new result which generalizes,
to curved manifolds, previous works in Minkowski spacetime. As a particular case, we have
analyzed the massless limit of the effective Lagrangian for both scalar and spinor
fields.

We have found, in four dimensions, that infrared divergences appear in the massless spinor case which
are purely caused by the gravitational field. This effect can be interpreted as follows:
The vacuum energy of charged spinors with small mass (or equivalently massive charged spinors for
which $M^{2}\ll B$) dramatically increases due to the presence of the gravitational field.
We expect that infrared divergences would manifest themselves also for massless scalar fields at
higher orders in the Riemannian curvature.




\acknowledgments
We would like to thank Prof. Ivan G. Avramidi for many useful discussions and for valuable
suggestions on how to improve the present manuscript.


\begin{thebibliography}{99}

\bibitem{avramidi94} Avramidi I G, (1994)
Covariant methods for calculating the low-energy
effective action in quantum field theory and quantum gravity,
arXiv:gr-qc/9403036, 48 pp

\bibitem{avramidi97} Avramidi I G, (1997)
Covariant approximation schemes for calculation of the heat kernel in quantum
field theory, in: \emph{Quantum Gravity}, Eds. V. A. Berezin, V. A. Rubakov and D. V.
Semikoz (Singapore: World Scientific), pp. 61-78

\bibitem{avramidi99} Avramidi I G,
A model of stable chromomagnetic vacuum in higher-dimensional Yang-Mills theory,
\emph{Prog. Phys.} {\bf 47} (1999) 433-455;


\bibitem{avramidi00} Avramidi I G, (2000)
Heat Kernel and Quantum Gravity, Lecture Notes in Physics, Series
Monographs, LNP: m64 (Berlin: Springer-Verlag)

\bibitem{avramidi02}
Avramidi I G,
Heat kernel approach in quantum field theory,
\emph{Nucl. Phys. Proc. Suppl.} {\bf 104} (2002) 3-32

\bibitem{avramidi08}
Avramidi I G,
Heat kernel on homogeneous bundles,
\emph{Int. J. Geom. Meth. Mod. Phys.}, {\bf 5} (2008) 1-23

\bibitem{avramidi08a} Avramidi I G,
Heat kernel on homogeneous bundles over symmetric spaces,
\emph{Comm. Math. Phys.} {\bf 288} (2009) 963-1006

\bibitem{avramidi08e}
Avramidi I G and Fucci G,
Nonperturbative heat kernel asymptotics on homogeneous Abelian bundles,
\emph{Comm. Math. Phys.} (2009) DOI: 10.1007/s00220-009-0804-6

\bibitem{avramidi09a}
Avramidi I G and Fucci G, (2009)
Low-energy effective action in non-perturbative electrodynamics in curved spacetime,
arXiv:0902.1541 [hep-th], 41pp


\bibitem{barvinsky85} Barvinsky A O and Vilkovisky G A,
The generalized Schwinger-DeWitt technique in gauge theories and quantum gravity,
\emph{Phys. Rep.} {\bf 119}, No. 1, (1985) 1-74

\bibitem{bastianelli09} Bastianelli F, Davila J M and Schubert C,
Gravitational corrections to the Euler-Heisenberg lagrangian
\emph{JHEP} 03 (2009) 086

\bibitem{cognola93} Cognola G, Kirsten K and Odintsov S D,
One-loop effective potential on hyperbolic manifolds
\emph{Phys. Rev.} D {\bf 48} (1993) 790-799

\bibitem{cognola94} Cognola G,
Renormalization of the one-loop effective action on an arbitrary curved space-time: A general method
\emph{Phys. Rev.} D {\bf 50} (1994) 909-916

\bibitem{dewitt65} DeWitt B S, (1965)
Dynamical Theory of Groups and Fields,
(Gordon and Breach Science Publishers)

\bibitem{dewitt67a} DeWitt B S,
Quantum theory of gravity II: The manifestly covariant theory,
\emph{Phys. Rev} {\bf 162} (1967) 1195-1238

\bibitem{dewitt67b} DeWitt B S,
Quantum theory of gravity III: The application of the covariant theory,
\emph{Phys. Rev} {\bf 162} (1967) 1239-1256

\bibitem{dewitt75} DeWitt B S,
Quantum field theory in curved spacetime,
\emph{Phys. Rep.} {\bf 19}, No. 6 (1975) 295-357

\bibitem{dewitt03}
DeWitt B S, (2003)
The Global Approach to Quantum Field Theory.
Oxford University Press, Oxford

\bibitem{dunne99} Dunne G V and Hall T M,
Borel summation of the derivative expansion and effective actions,
\emph{Phys. Rev.} D {\bf 60} (1999) 065002

\bibitem{dunne04} Dunne G V, Heisenber-Euler effective Lagrangians: Basics and extensions, in
Ian Kogan Memorial Collection, \emph{From Fields to Strings: Circumnavigating Theoretical Physics}, Vol. 1,
M. A. Shifman et al. (Eds.), World Scientific, Singapore, (2004), 445

\bibitem{dunne05} Dunne G V and Schubert C,
Worldline instantons and pair production in inhomogeneous fields,
\emph{Phys. Rev.} D {\bf 72} (2005) 105004

\bibitem{elizalde94} Elizalde E, Odintsov S D, Romeo A, Bytsenko A and Zerbini S, (1994)
Zeta Regularization Techniques with Applications (World Scientific, Singapore)

\bibitem{erdelyi53}  Erd\'{e}lyi A, (1953)
Higher Transcendental Functions, Vol. 1,
Bateman Project Staff (New York, McGraw-Hill)

\bibitem{finch03} Finch S R, (2003) Mathematical Constants,
in \emph{Encyclopedia Math. Appl.} vol. {\bf 94},
Cambridge University Press, Cambridge


\bibitem{gilkey95}
Gilkey P B, (1995)
Invariance Theory, the Heat Equation and the
Atiyah-Singer Index Theorem, (Boca Raton: CRC Press)

\bibitem{gradshtein07}  Gradshtein I S and Ryzhik I M, (2007)
Table of Integrals, Series and Products, Eds. A. Jeffrey and D.
Zwillinger (Oxford: Academic)

\bibitem{hawking77} Hawking S W,
Zeta function regularization of path integrals in curved spacetime
\emph{Comm. Math. Phys.} {\bf 55} (1977) 133-148

\bibitem{miller98} Miller J and Adamchik V S,
Derivatives of the Hurwitz zeta function for
rational arguments,
\emph{J. Comput. Appl. Math.} {\bf 100} (1998) 201-206

\bibitem{kirsten01}
Kirsten K, (2001)
Spectral Functions in Mathematics and Physics,
(Boca Raton: CRC Press)

\bibitem{schwinger54} Schwinger  J S,
The theory of quantized fields V.,
\emph{Phys. Rev.} {\bf 93} (1954) 615-628

\bibitem{schwinger51}
Schwinger J S
On gauge invariance and vacuum polarization,
\emph{Phys. Rev.} {\bf 82} (1951) 664--679


\bibitem{soldati98} Soldati R and Sorbo L,
Effective action for Dirac spinors in the presence of general uniform electromagnetic fields,
\emph{Phys. Lett.} {\bf B426} (1998) 82-88

\bibitem{vassile03} Vassilevich  D V,
Heat kernel expansion: User's manual,
\emph{Phys. Rep.} {\bf 388} (2003) 279-360


\end{thebibliography}
\end{document}